\documentclass{appolb}

 \usepackage[dvips,final]{graphicx}
  \usepackage{amssymb}
   \usepackage{amsmath}
    \usepackage{amsfonts}
     \usepackage{epsfig}
      \usepackage{bm}

\begin{document}
\title{Charm meson production and double parton interactions\\ at the LHC %
\thanks{Presented at the XX Cracow EPIPHANY Conference on the Physics at the LHC}%
}
\author{Rafa{\l} Maciu{\l}a, A.~van~Hameren
\address{Institute of Nuclear Physics PAN, PL-31-342 Cracow, Poland}
\\
{Antoni Szczurek}
\address{Institute of Nuclear Physics PAN, PL-31-342 Cracow, Poland and\\
University of Rzesz\'ow, PL-35-959 Rzesz\'ow, Poland}
}
\maketitle
\begin{abstract}
We discuss production of open charm mesons in proton-proton collisions at the LHC. The cross section for inclusive production of $c \bar c$ pairs is calculated within the $k_{\perp}$-factorization approach in which a major part of higher-order corrections is belived to be effectively included. We use and test the applicability of several unintegrated gluon distributions. Numerical results of the $k_{\perp}$-factorization approach are compared to NLO pQCD collinear predictions. Inclusive differential distributions in transverse momentum and (pseudo)rapidity of several charmed mesons are presented and compared to recent results of the ALICE, ATLAS
and LHCb collaborations. We also examine production of neutral charmed meson-antimeson pairs 
($D^0 {\overline D^0}$) in unique kinematics of forward rapidities of the LHCb experiment. Kinematical correlations in azimuthal angle $\varphi_{D^0 {\overline D^0}}$, invariant mass $M_{D^0 {\overline D^0}}$ and rapidity difference $Y_{D^0 {\overline D^0}}$ distributions are calculated. Moreover, we also discuss production of two pairs of $c \bar c$ within a simple
formalism of double-parton scattering (DPS).
We compare results of calculations of single-parton scattering (SPS) 
and double-parton scattering (DPS) for production of 
$c \bar c c \bar c$ and for $D^0-D^0$ meson-meson correlations.
We compare our predictions for double charm production with recent results of
the LHCb collaboration for azimuthal angle $\varphi_{D^0 D^0}$ distribution, dimeson invariant mass $M_{D^0 D^0}$ and rapidity distance between
mesons $Y_{D^0 D^0}$.
The obtained results clearly certify the dominance of DPS in the
production of events with double charm.
\end{abstract}
  
\section{Introduction}

Recently, ATLAS \cite{ATLASincD}, ALICE \cite{ALICEincD,ALICEincDs} 
and LHCb \cite{LHCbincD} collaborations have measured inclusive 
distributions of different charmed mesons. The LHCb collaboration has measured in addition  
a few correlation observables for charmed meson-antimeson pairs in the forward rapidity region $2 < y < 4$ \cite{LHCb-DPS-2012}.
Previously, the STAR collaboration at RHIC has measured only $e-D$ correlation of charmed mesons
and leptons from their semileptonic decays \cite{Mischke}. An examination of $D\overline D$ meson-antimeson
correlations was accessible only at the Tevatron where first midrapidity measurements
of $D\overline D$ azimuthal angle correlations have been performed by the CDF experiment \cite{Tevatron-DD}.

Commonly in the exploration of heavy quark production the main efforts concentrate on inclusive distributions.
Improved schemes of standrad pQCD NLO collinear approach, e.g. FONLL \cite{FONLL} or GM-VFNS \cite{GM-VFNS} are state of art in this
respect. These models can be, however, used  only when transverse
momenta of charm quark and antiquark are balanced. This means in practice 
that it cannot be used for studies of correlation observables, which provide broader kinematical characteristic
of the process under consideration.

Another approach which is much more efficient in this respect is the so-called $k_{\perp}$-factorization (see e.g.~\cite{Maciula:2013wg} and references therein). Here, the transverse momenta of incident partons are explicitly taken into account and their emission is encoded
in the unintegrated gluon distributions -- the building blocks of the formalism. This allows to construct different correlation distributions which are strictly related with the transverse momenta of initial particles. 

In addition, within the $k_{\perp}$-factorization approach it is possible to study interesting low-x effects, which may appear especially in the case of charm production. In principle a comparison of experimental data and predictions with the unintegrated gluon distribution functions (UGDFs) which include such 
effects may tell us more, e.g. about indication of the saturation -- the topic being extensively studied in recent years.

Moreover, it was recently argued that the cross section
for $c \bar c c \bar c$ production at LHC energies may be very large
due to mechanism of double-parton scattering (DPS), which is a completely new situation \cite{Luszczak:2011zp,Cazaroto:2013fua}. 
The double scattering effects were studied in several other processes
such as four jet production, production of $W^+ W^-$ pairs or production of four charged leptons, however, 
in all the cases the DPS contributions have been found to be much smaller than
the conventional single-parton scattering (SPS) mechanisms. 

In the meanwhile the LHCb collaboration measured the
cross section for the production of $D D$ meson-meson pairs at 
$\sqrt{s}$ = 7 TeV which is surprisingly large, including interesting correlation distributions \cite{LHCb-DPS-2012}.
So far those data sets for double open charm production have been studied differentially only
within the $k_{\perp}$-factorization approach using unintegrated gluon distributions \cite{Maciula:2013kd}, where several observables useful to identify the DPS effects in the case of double open charm production have been carefully discussed.

Separately the production of double hidden charm was studied e.g. in
Ref.~\cite{Baranov:2012re} for the $p p \to J/\psi J/\psi X$ process.
There the SPS single-$J/\psi$ and DPS double-$J/\psi$ contributions are comparable. Furthermore, the DPS contribution
exceeds the SPS contribution for large rapidity distance between
the two $J/\psi$'s. This is similar to the case of $c \bar c c \bar c$ production \cite{Maciula:2013kd}.

In order to draw definite conclusions about the DPS effects in double-$D$ meson production it is necessary
to carefully estimate contribution to $c \bar c c \bar c$ final state from the standard mechanism of single-parton scattering.
The latter mechanism constitutes higher-order correction to conventional SPS single $c \bar c$ production and one may expect suppression
in comparison to the DPS contribution, however, it should be accurately calculated in order to reduce uncertainty of the DPS theoretcal model. So far the SPS contribution was calculated only in high-energy approximation \cite{Schafer:2012tf}, which is relevant for large rapidity separation between produced mesons. Since, in the LHCb experiment the condition of large rapidity distances is not always fulfilled, it seems to be essential to perform exact calculations.

\section{Inclusive single charm production}

As discussed in Ref.~\cite{Maciula:2013wg}, in the case of charm production in proton-proton scatering at the LHC one enters a quite new kinematical and dynamical domain. Since the production of charm quarks at high energies
is known to be dominated by the gluon-gluon fusion, the charm production at the LHC
can be used to verify the quite different models of UGDFs. It is very interesting and important
to test various unintegrated gluon distributions from the literature \cite{KMR,KMS,Kutak-Stasto,Jung} in hadronic reactions, in the kinematical regimes never achieved before. In contrast to the collinear gluon distributions (PDFs) the UGDFs are based on different theoretical assumptions and differ considerably among themselves. 
Therefore, one may expect that they could lead to different production rates of $c \bar c$ pairs at the LHC. 

The cross section for the production of a pair of charm quark -- 
charm antiquark can be written as:
\begin{eqnarray}
\frac{d \sigma(p p \to c \bar c X)}{d y_1 d y_2 d^2 p_{1t} d^2 p_{2t}} 
&& = \frac{1}{16 \pi^2 {\hat s}^2} \int \frac{d^2 k_{1t}}{\pi} \frac{d^2 k_{2t}}{\pi} \overline{|{\cal M}^{off}_{g^{*}g^{*} \to c\; \bar c}|^2} \nonumber \\
&& \times \;\; \delta^2 \left( \vec{k}_{1t} + \vec{k}_{2t} - \vec{p}_{1t} - \vec{p}_{2t}
\right)
{\cal F}_g(x_1,k_{1t}^2,\mu^2) {\cal F}_g(x_2,k_{2t}^2,\mu^2).\nonumber \\
\end{eqnarray}
The main ingredients in the formula are off-shell matrix element for $g^{*}g^{*} \rightarrow c \;\bar{c}$ subprocess
and unintegrated gluon distributions (UGDF). The relevant matrix
elements are known and can be found e.g. in Ref.~\cite{CCH91}. 
The unintegrated gluon distributions are functions of
longitudinal momentum fraction $x_1$ or $x_2$ of gluon with respect to its parent nucleon and of gluon transverse momenta $k_{t}$.
Some of them depend in addition on the
factorization scale $\mu$.
The longitudinal momentum fractions can be calculated as:
\begin{eqnarray}
x_1 = \frac{m_{1t}}{\sqrt{s}}\exp( y_1) 
     + \frac{m_{2t}}{\sqrt{s}}\exp( y_2),\nonumber \\
x_2 = \frac{m_{1t}}{\sqrt{s}}\exp(-y_1)
     + \frac{m_{2t}}{\sqrt{s}}\exp(-y_2),
\end{eqnarray}
where $m_{it} = \sqrt{p_{it}^2 + m_Q^2}$ is the transverse mass of produced quark/antiquark.

The numerical quark-level results may be compared to real experimental data after inclusion of the hadronization effects. The transition from the quark level to open heavy meson states has to be performed. In the case of charm (or bottom) particles, the hadronization is
usually done with the help of fragmentation functions. The inclusive distributions of
charmed mesons can be then obtained through a convolution of inclusive distributions
of charm quarks/antiquarks and $c \to $ D fragmentation functions:
\begin{equation}
\frac{d \sigma(pp \rightarrow D \bar{D} X)}{d y_D d^2 p_{t,D}} \approx
\int_0^1 \frac{dz}{z^2} D_{c \to D}(z)
\frac{d \sigma(pp \rightarrow c \bar{c} X)}{d y_c d^2 p_{t,c}}
\Bigg\vert_{y_c = y_D \atop p_{t,c} = p_{t,D}/z}
 \; ,
\label{Q_to_h}
\end{equation}
where $p_{t,c} = \frac{p_{t,D}}{z}$ and $z$ is the fraction of
longitudinal momentum of heavy quark carried by meson.
We have made typical approximation assuming that $y_{c}$  is
unchanged in the fragmentation process, i.e. $y_D = y_c$.

In our calculations we use standard Peterson model of fragmentation function \cite{Peterson} with
the parameter $\varepsilon_{c} = 0.02$ for pseudoscalar, and BCFY model \cite{BCFY} with $r_{c}=0.1$ for vector $D$ meson states, respectively. 
This is consistent with the fragmentation scheme applied in the FONLL framework, where rather hard fragmentation functions for charm quarks are suggested \cite{Cacciari-RHIC}. This issue as well as effects of applying other fragmentation functions from the literature, together with aspects of QCD evolution, are carefully discussed in Ref.~\cite{Maciula:2013wg}. The fragmentation functions used here are normalized to branching fractions BR($c \to D$) from Ref.~\cite{Lohrmann2011}.

\subsection{ALICE}

\begin{figure}[!h]
\begin{minipage}{0.47\textwidth}
 \centerline{\includegraphics[width=1.0\textwidth]{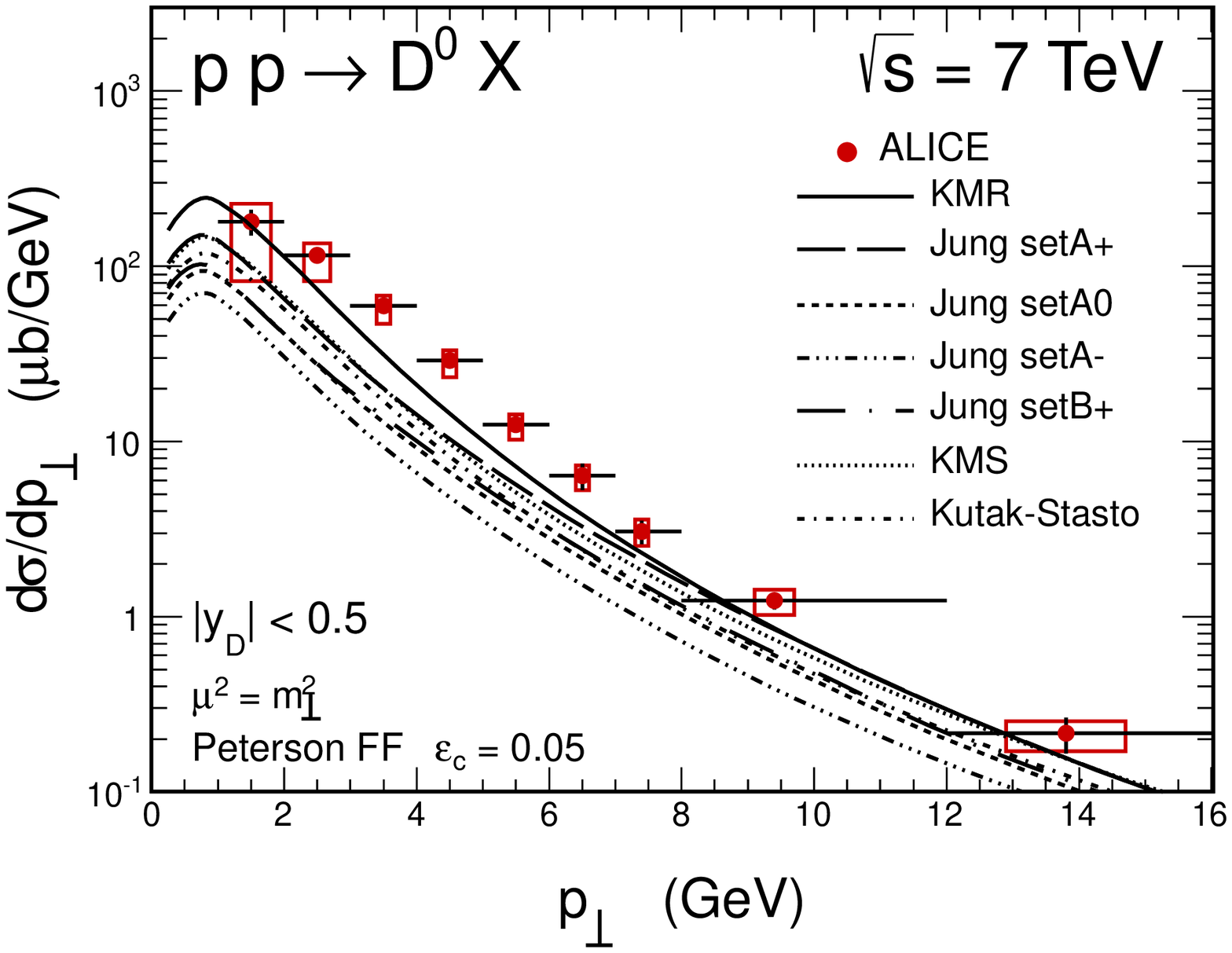}}
\end{minipage}
\hspace{0.5cm}
\begin{minipage}{0.47\textwidth}
 \centerline{\includegraphics[width=1.0\textwidth]{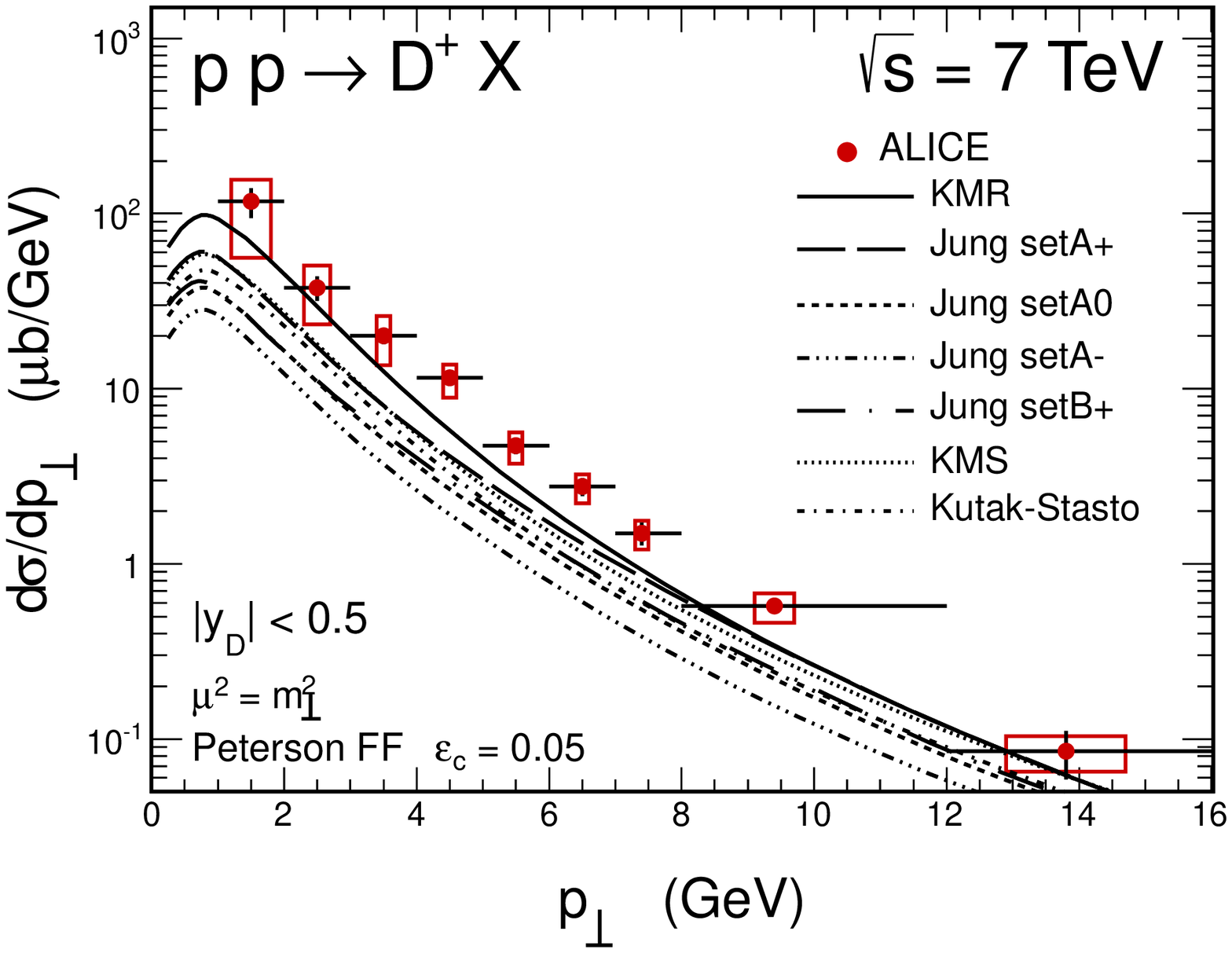}}
\end{minipage}
   \caption{
\small Transverse momentum distribution of $D^0$ (left) and $D^{+}$ (right) mesons for different UGDFs together with the ALICE data. Details of the calculations are specified in the figure.}
 \label{fig:pt-alice-D-1}
\end{figure}

The ALICE collaboration has measured the transverse
momentum distribution of $D^0$, $D^+$, $D^{*+}$, $D_s^+$ mesons \cite{ALICEincD,ALICEincDs}. 
In the very limited range of rapidity $|y|< 0.5$ one tests unintegrated gluon distributions in a
pretty narrow region of longitudinal momentum fractions $10^{-4} \lesssim x \lesssim 10^{-2} $ \cite{Maciula:2013wg}.
In Fig.~\ref{fig:pt-alice-D-1} we present transverse momentum distribution
of $D^0$ (left panel) and $D^{+}$ (right panel) mesons . We show results for different UGDF
known from the literature. Most of the applied unintegrated distributions fail
to describe the ALICE data. Only the KMR UGDF provides the results which are
close to the measured distributions.

\begin{figure}[!h]
\begin{minipage}{0.47\textwidth}
 \centerline{\includegraphics[width=1.0\textwidth]{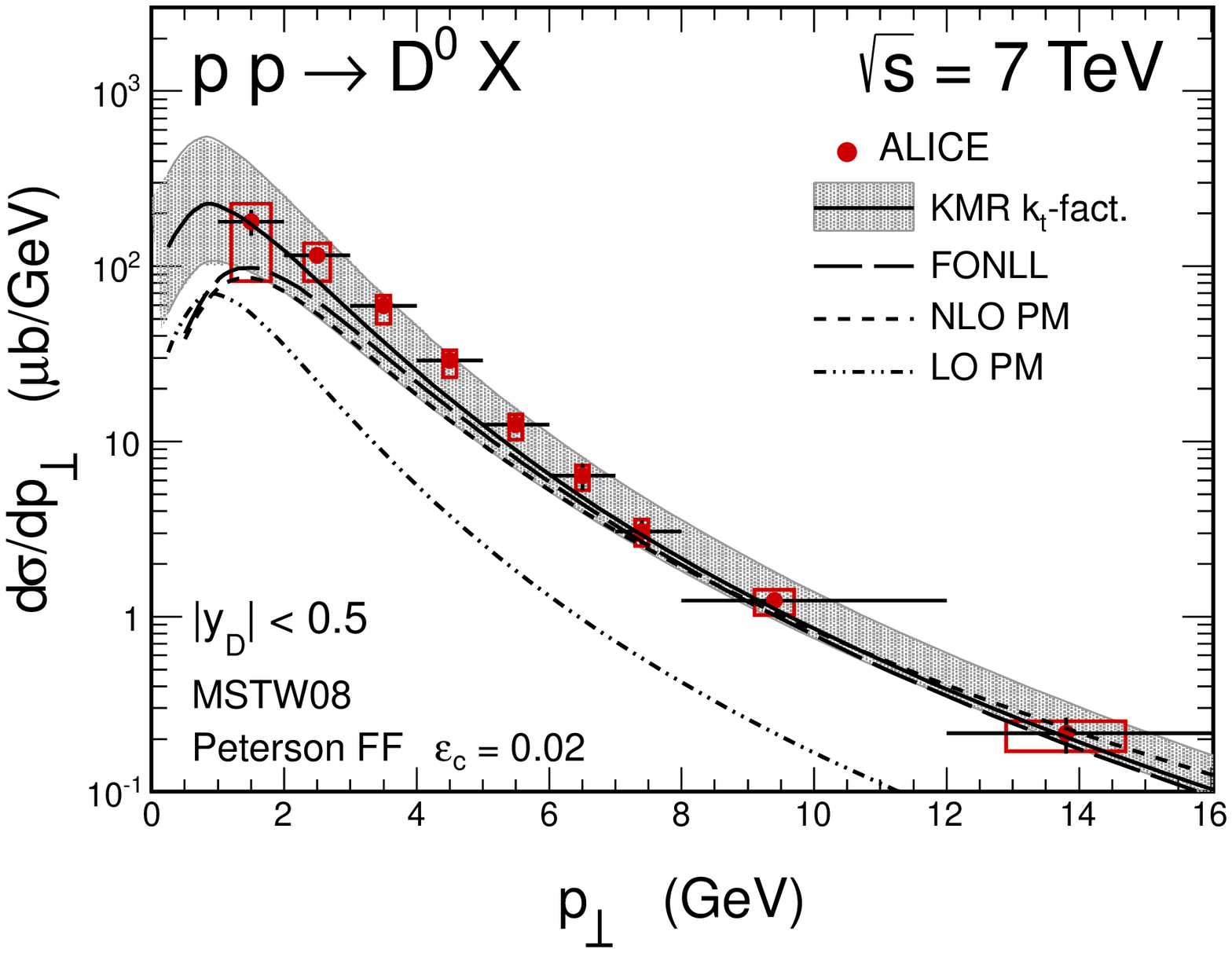}}
\end{minipage}
\hspace{0.5cm}
\begin{minipage}{0.47\textwidth}
 \centerline{\includegraphics[width=1.0\textwidth]{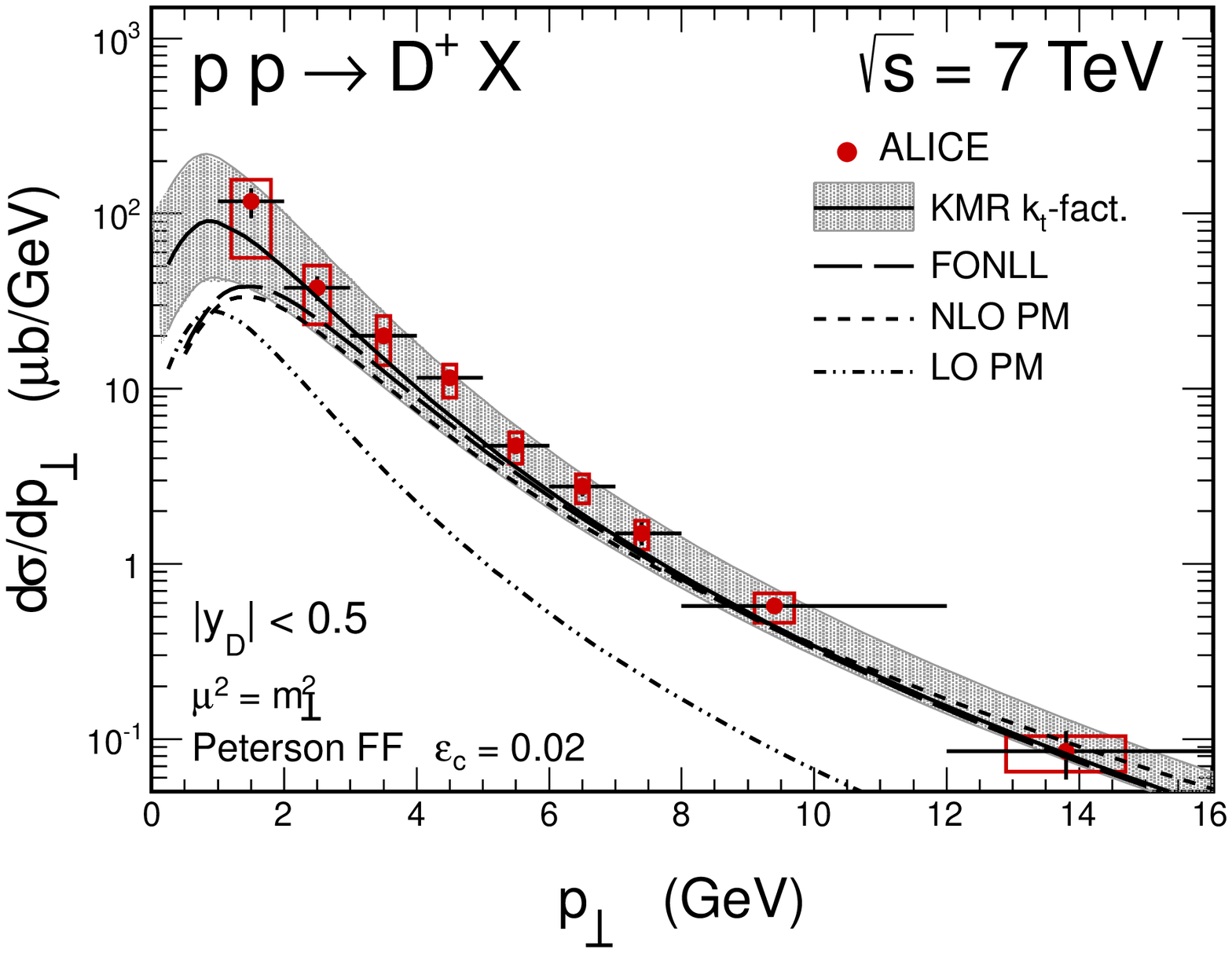}}
\end{minipage}
   \caption{
\small Transverse momentum distribution of $D^0$ (left) and $D^{+}$ (right) mesons for the ALICE
kinematical region. Together with our predictions for the KMR UGDF (solid line with shaded band) results of different other popular approaches are also shown.}
 \label{fig:pt-alice-D-5}
\end{figure}

In Fig.~\ref{fig:pt-alice-D-5} we present
a comparison of our calculations to the results of some other
popular approaches used in the literature. Our results obtained within the $k_{\perp}$-factorization
approach with the KMR UGDF are very similar to those obtained within
NLO PM and FONLL models. The cross sections obtained within
leading-order collinear approximation (LO PM) are much smaller, in particular
for larger transverse momenta. In this case the uncertainties coming from the perturbative part
of the calculation are also drawn. The uncertainties of our predictions are obtained by
changing charm quark mass $m_c = 1.5 \pm 0.3$ GeV and by varying renormalization
and factorization scales $\mu^{2} = \zeta m_{t}^{2}$, where $\zeta \in (0.5; 2)$. The gray shaded bands
represent these both sources of uncertainties summed in quadrature.

\subsection{ATLAS}

The ATLAS experiment covers much broader range of pseudorapidities than ALICE.
As a consequence one tests a bit wider region of longitudinal momentum fractions.
However, the gluon distributions in this range of $x$ values carried by gluons 
are also rather well known, so the application of the known UGDFs should be reliable too.

\begin{figure}[!h]
\begin{minipage}{0.47\textwidth}
 \centerline{\includegraphics[width=1.0\textwidth]{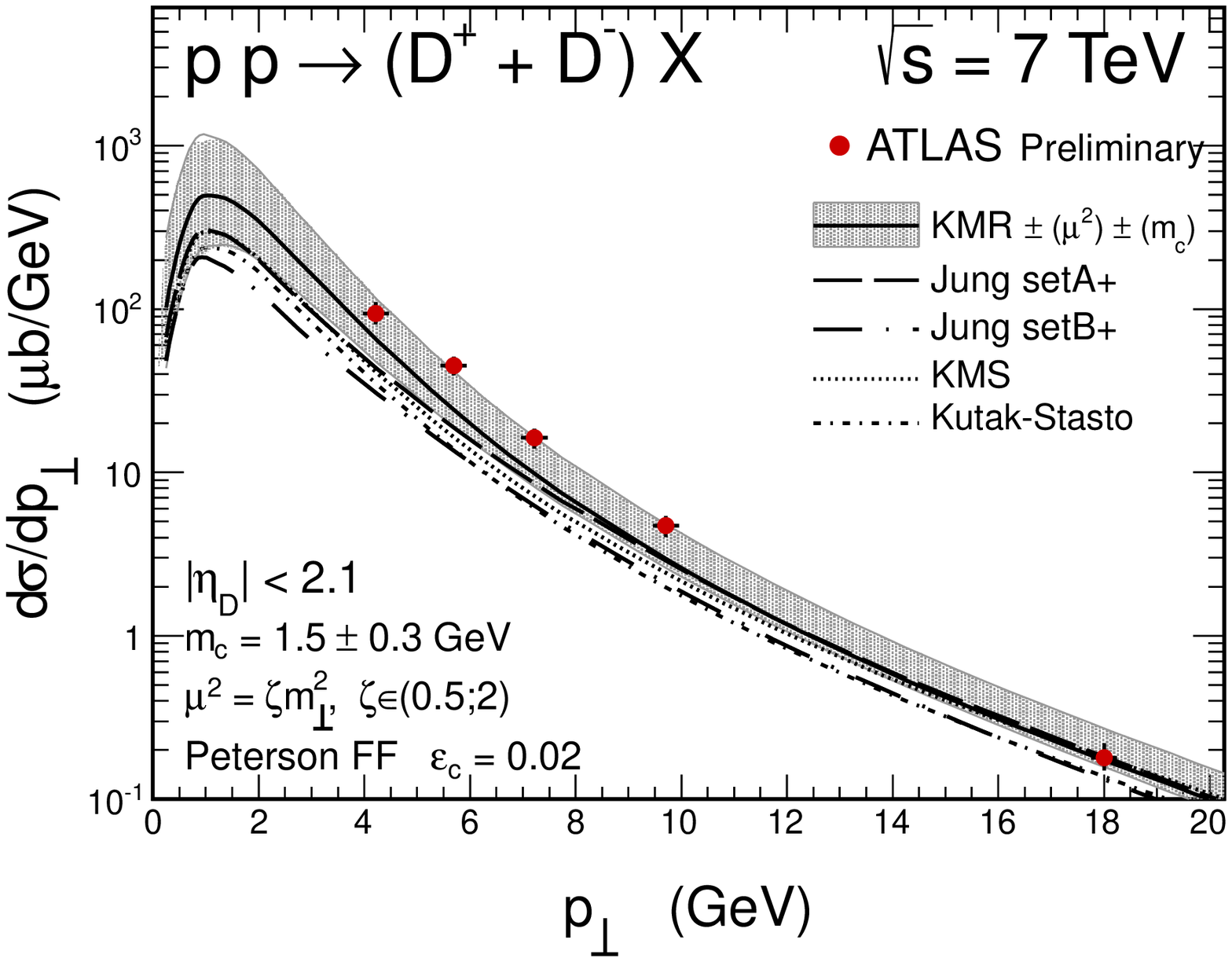}}
\end{minipage}
\hspace{0.5cm}
\begin{minipage}{0.47\textwidth}
 \centerline{\includegraphics[width=1.0\textwidth]{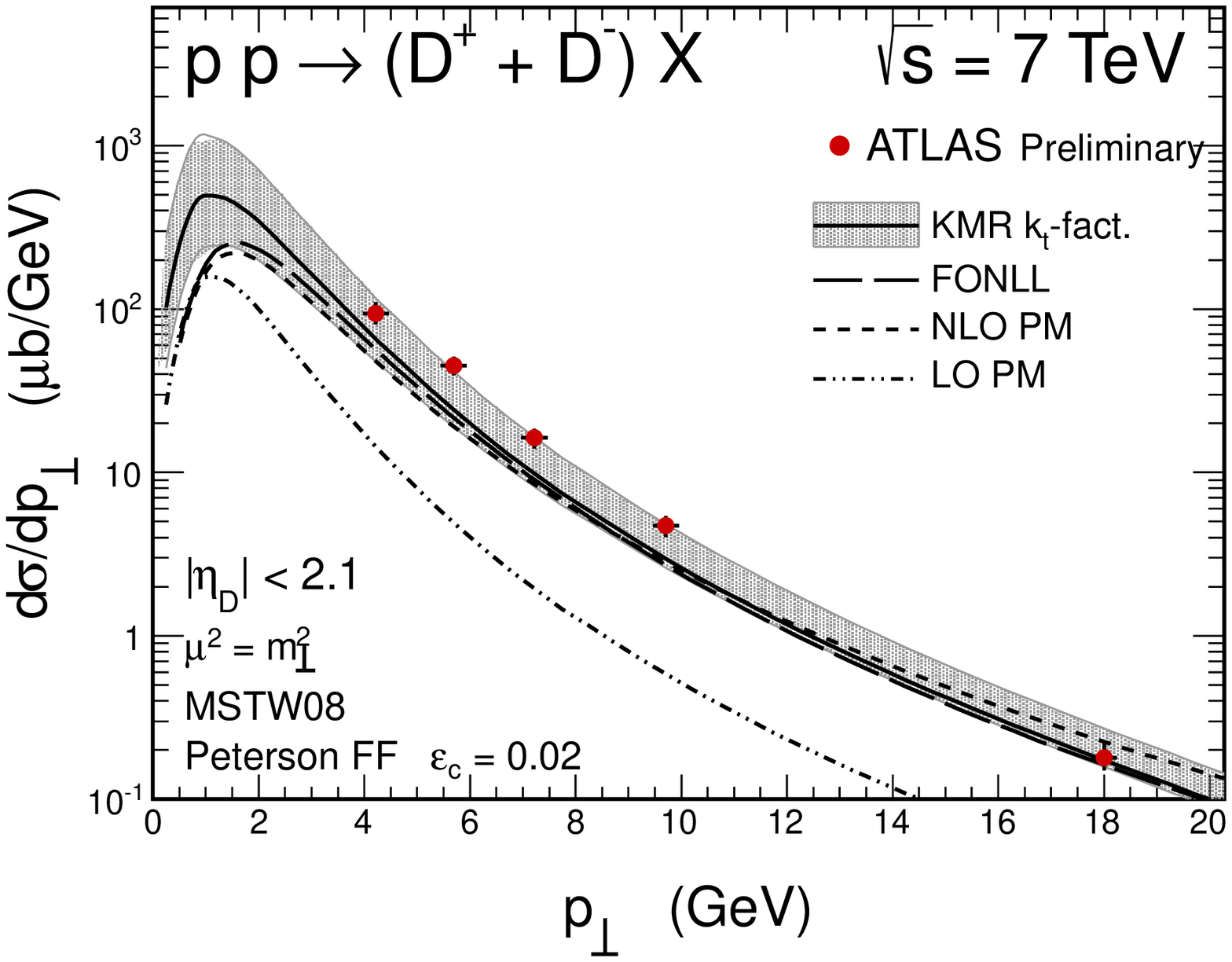}}
\end{minipage}
   \caption{
\small Transverse momentum distribution of $D^{\pm}$ mesons for different UGDFs (left) and for
standard approaches (right) compared with the ATLAS experimental data.}
 \label{fig:pt-atlas-D-2a}
\end{figure}

The left panel of Fig.~\ref{fig:pt-atlas-D-2a} presents transverse momentum distributions of charged pseudoscalar $D^{\pm}$ mesons for different models of unintegrated gluon distributions. Overall situation is very similar as for the ALICE experiment except of the agreement with the experimental data points, which is somewhat worse in this case. Only the very upper limit of the KMR result is consistent
with the ATLAS data. This may be caused by much broader range of pseudorapidities in the case of the ATLAS detector. Potentially, this can be related to double-parton scattering effects \cite{Maciula:2013kd}. The other standard pQCD approaches also give results below the ATLAS data as can be seen in the right panel of Fig.~\ref{fig:pt-atlas-D-2a}. 

\begin{figure}[!h]
\begin{minipage}{0.47\textwidth}
 \centerline{\includegraphics[width=1.0\textwidth]{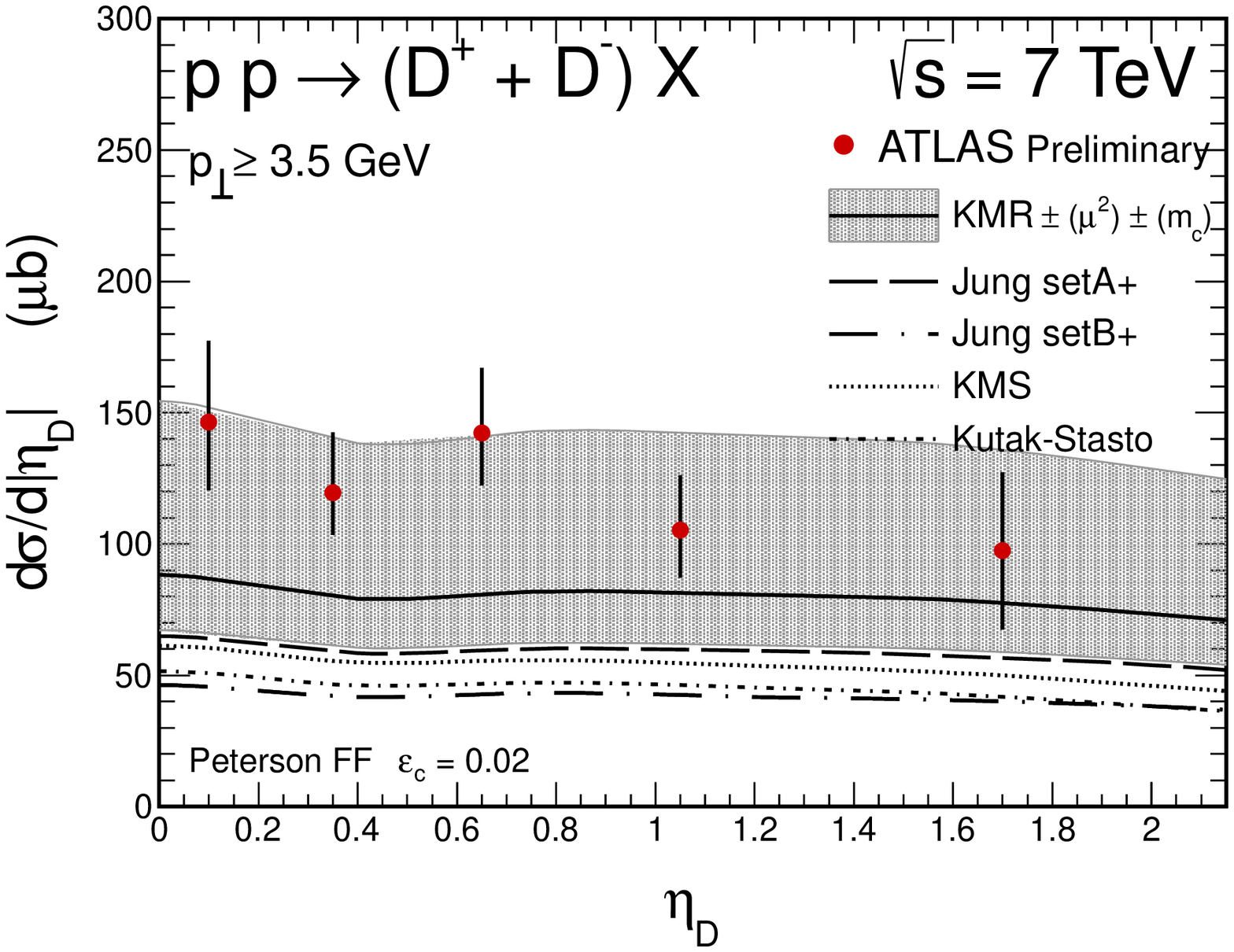}}
\end{minipage}
\hspace{0.5cm}
\begin{minipage}{0.47\textwidth}
 \centerline{\includegraphics[width=1.0\textwidth]{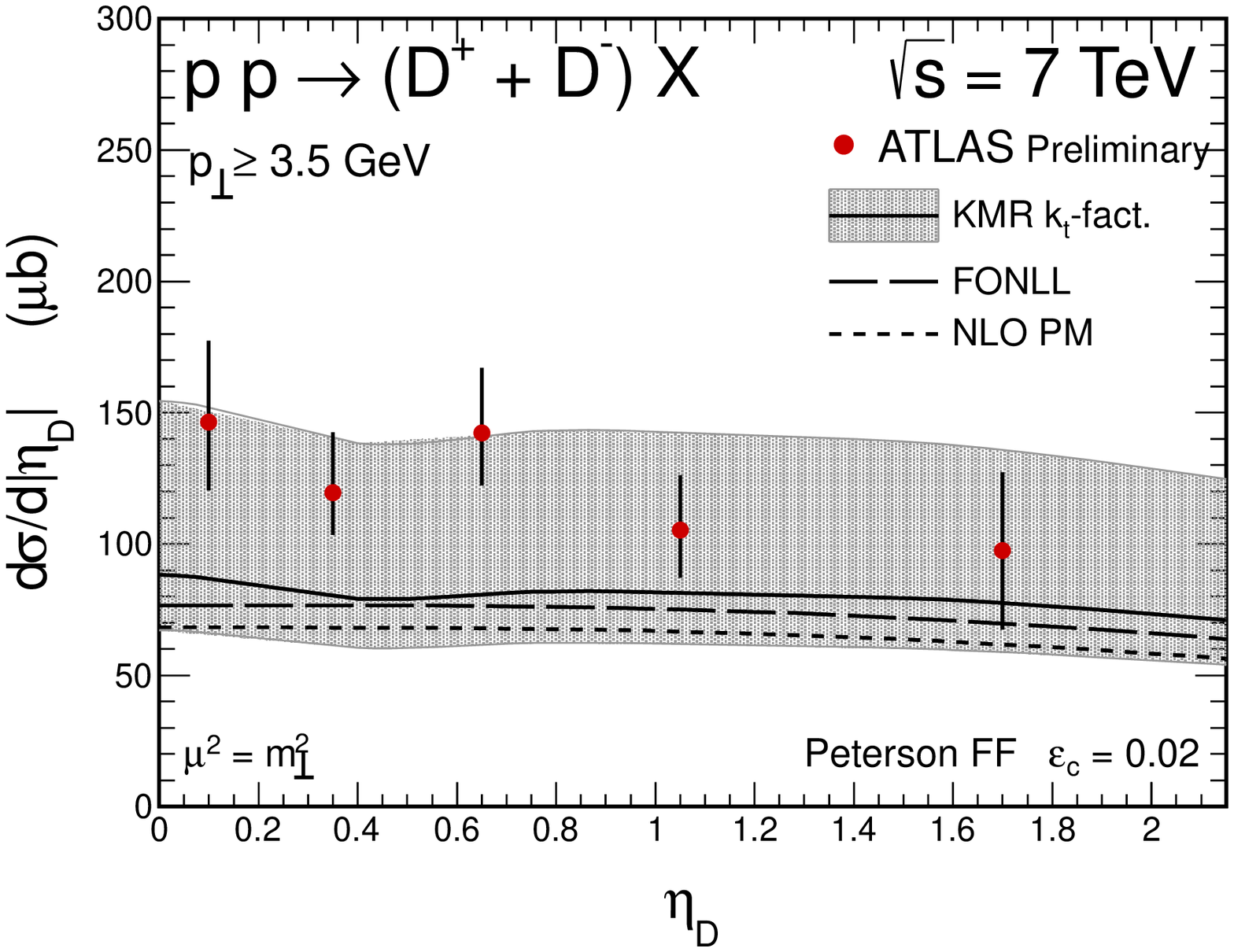}}
\end{minipage}
   \caption{
\small Distribution in $D^{\pm}$ meson pseudorapidity. The results for
different UGDFs (left) sa well as for other standard approaches (right) are compared with the ATLAS data.
}
 \label{fig:eta-atlas-D-1b}
\end{figure}

Fairly large span of pseudorapidities allows the ATLAS collaboration to extract
also pseudorapidity distributions. In Fig.~\ref{fig:eta-atlas-D-1b} we show pseudorapidity distributions for charged $D^{\pm}$
meson. These distributions are rather flat. As in the case of the transverse momentum distributions, here also only the upper limits of large error bars of the theoretical results obtained with the KMR distributions are consistent
with the ATLAS data. The results with other UGDFs clearly underpredict the experimental points (left panel).
The central value of the $k_{\perp}$-factorization approach (grey band) with the KMR UGDF is consistent with the FONLL and NLO PM predictions (right panel).

\subsection{LHCb}

At the end we focus on unique measurements in the forward rapidity region $2 < y < 4.5$.
Recently the LHCb collaboration presented first results for
the production of $D^0$, $D^+$, $D^{*+}$ and $D_s^+$ mesons \cite{LHCbincD} in this region of phase space that has never been explored before.
In this case one can test asymmetric configuration of gluon longitudinal
momentum fractions: $x_1 \sim$ 10$^{-5}$ and $x_2 >$ 10$^{-2}$ \cite{Maciula:2013wg}.
Standard collinear gluon distributions as well as unintegrated one were never tested at such small values of $x_1$.
Moreover many models of the latter may be not good enough for $x_2 >$ 10$^{-2}$.
From this reason this is certainly more difficult region for reliable calculation and 
interpretation of experimental data and therefore special care in interpreting the results is required.

\begin{figure}[!h]
\begin{minipage}{0.47\textwidth}
 \centerline{\includegraphics[width=1.0\textwidth]{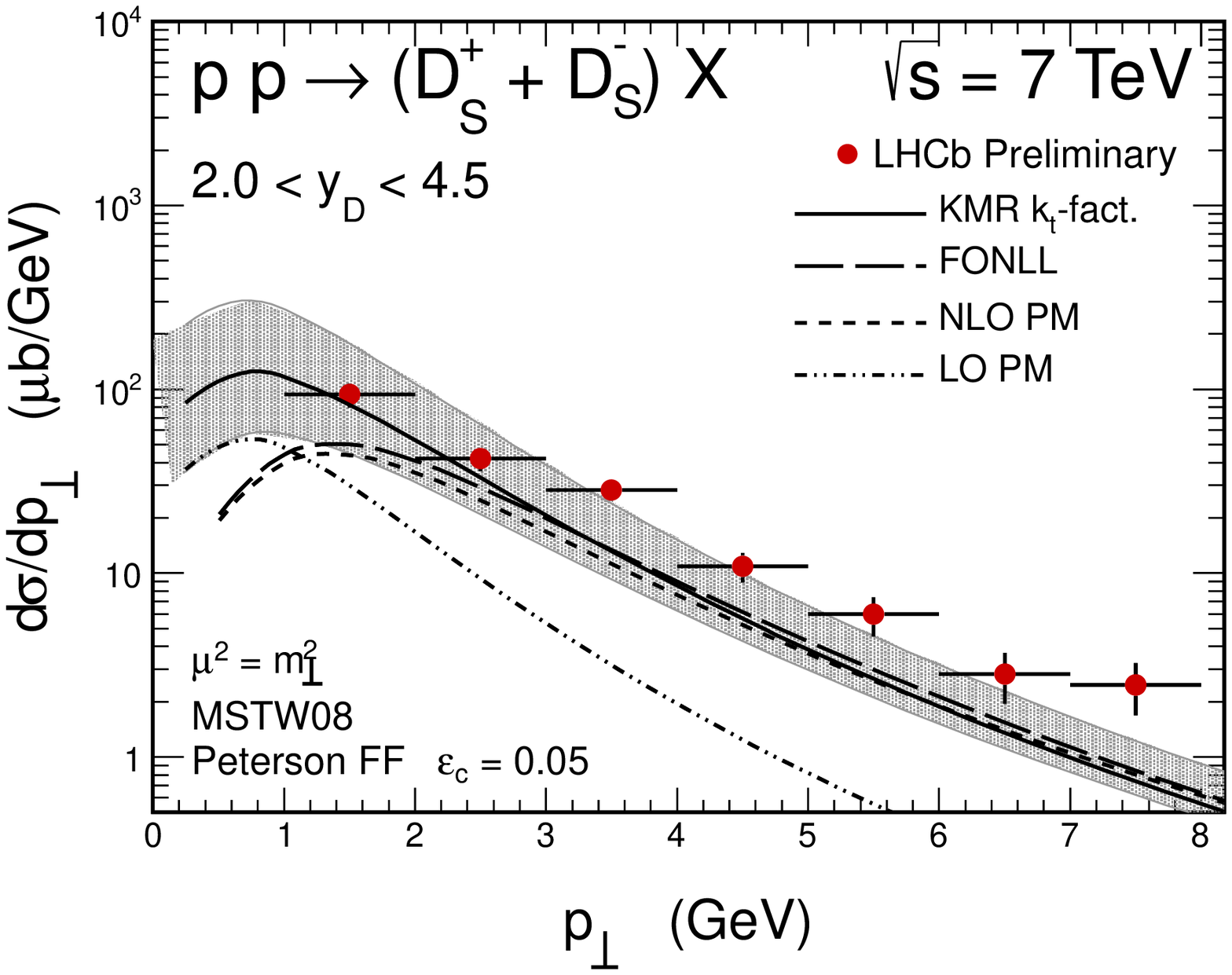}}
\end{minipage}
\hspace{0.5cm}
\begin{minipage}{0.47\textwidth}
 \centerline{\includegraphics[width=1.0\textwidth]{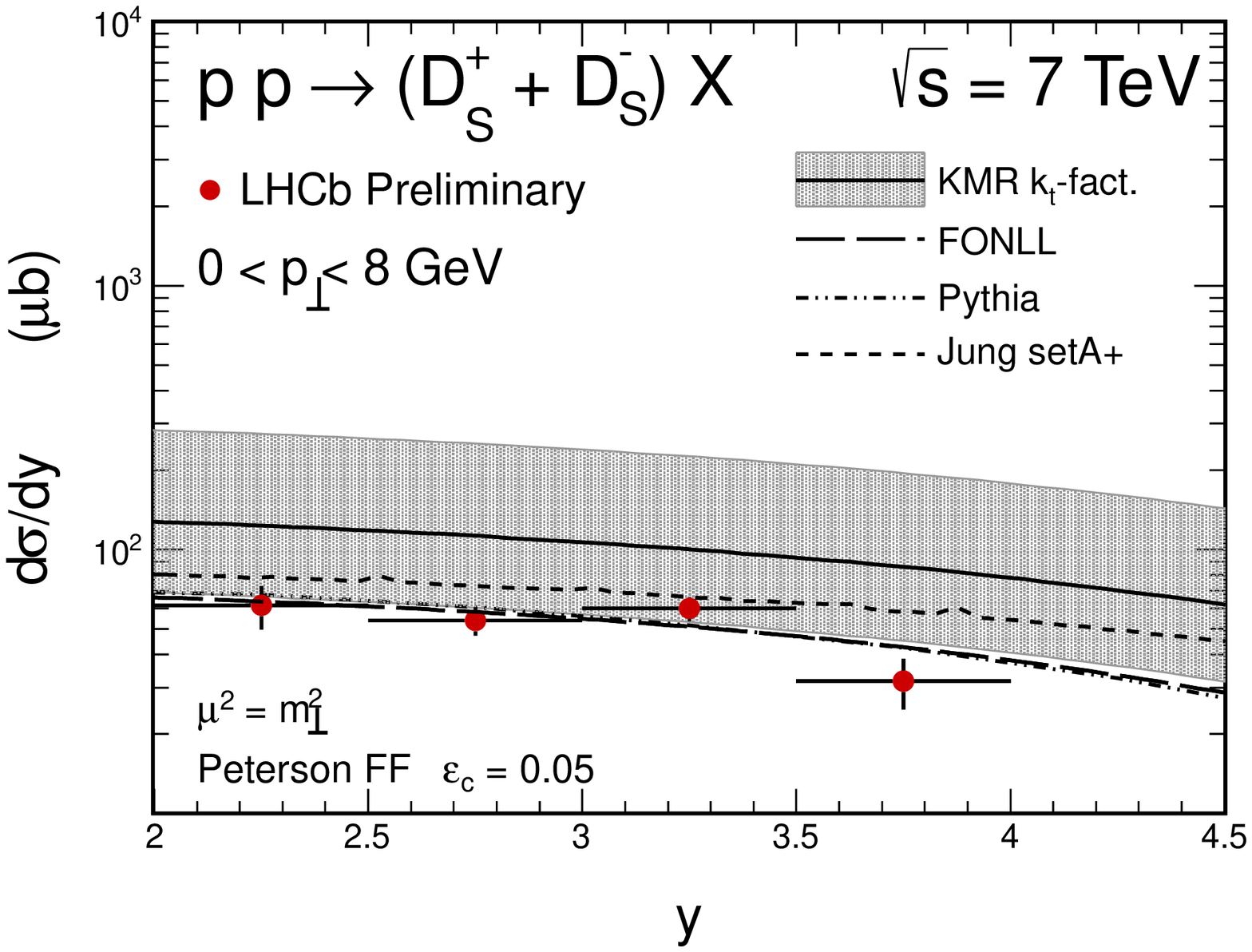}}
\end{minipage}
   \caption{
\small Results with overall uncertainties for transverse momentum (left) and rapidity distributions (right)
of $D_s^{\pm}$ for the $k_{\perp}$-factorization approach with the KMR UGDF.
For comparison we show predictions of other popular approaches.
}
 \label{fig:pt-lhcb-D-4}
\end{figure}

The LHCb, similar as ALICE, has measured also distributions of rather
rarely produced $D_s^{\pm}$ mesons.
In the left panel of Fig.~\ref{fig:pt-lhcb-D-4} we present transverse momentum distributions for $D_s^{\pm}$ mesons
distributions together with predictions of other popular approaches. In the right panel we show corresponding rapidity distribution calculated with different UGDFs and those obtained by applying other standard approaches.
The main conclusions are the same as for ALICE and ATLAS conditions. Our results with the KMR UGDF within uncertainties are consistent
with the experimental data and with the FONLL and NLO PM predictions.

\subsection{Charm-anticharm correlations at the LHCb}

In order to calculate correlation observables for $D\overline D$ pair production, measured recently in the LHCb experiment \cite{LHCb-DPS-2012}, we follow here, similar as in the single meson production,
the fragmentation function technique for hadronization process:
\begin{equation}
\frac{d \sigma(pp \to D \overline{D} X)}{d y_1 d y_{2} d^2 p_{1t}^{D} d^2 p_{2t}^{\overline{D}}}
 \approx
\int \frac{D_{c \to D}(z_{1})}{z_{1}}\cdot \frac{D_{\bar c \to \overline D}(z_{2})}{z_{2}}\cdot
\frac{d \sigma(pp \to c \bar{c} X)}{d y_1 d y_{2} d^2
  p_{1t}^{c} d^2 p_{2t}^{\bar{c}}} d z_{1} d z_{2} \; ,
\end{equation}
where: 
$p_{1t}^{c} = \frac{p_{1,t}^{D}}{z_{1}}$, $p_{2,t}^{\bar{c}} =
  \frac{p_{2t}^{\bar{D}}}{z_{2}}$ and
meson longitudinal fractions  $z_{1}, z_{2}\in (0,1)$.
The multidimensional distribution for $c$ quark and $\bar c$ antiquark is convoluted with respective
fragmentation functions simultaneously. As the
result of the hadronization one obtains corresponding two-meson 
multidimensional distribution. In the last step experimental kinematical
cuts on the distributions can be imposed. Then the resulting
distributions can be compared with experimental ones.

\begin{figure}[!h]
\begin{minipage}{0.47\textwidth}
 \centerline{\includegraphics[width=1.0\textwidth]{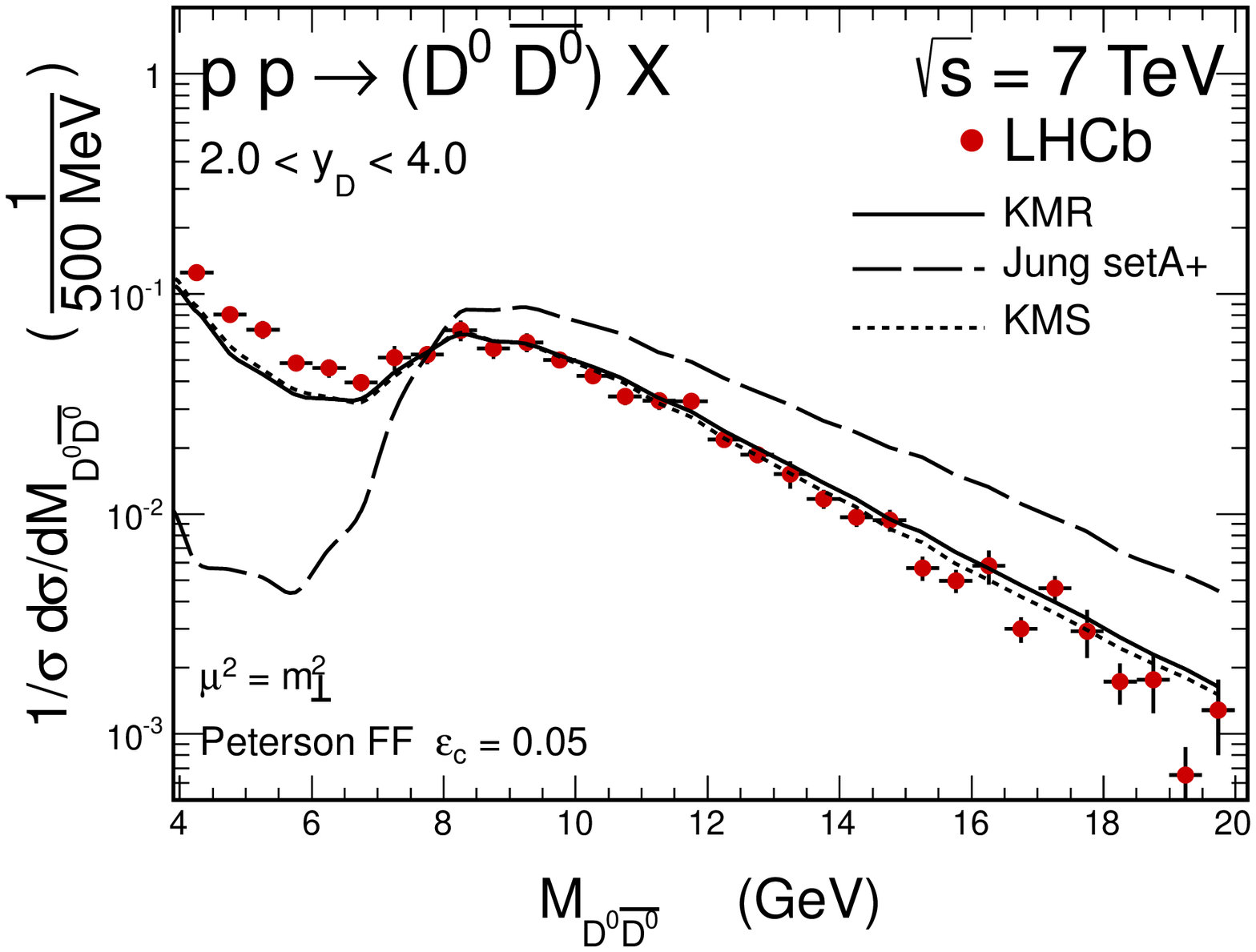}}
\end{minipage}
\hspace{0.5cm}
\begin{minipage}{0.47\textwidth}
 \centerline{\includegraphics[width=1.0\textwidth]{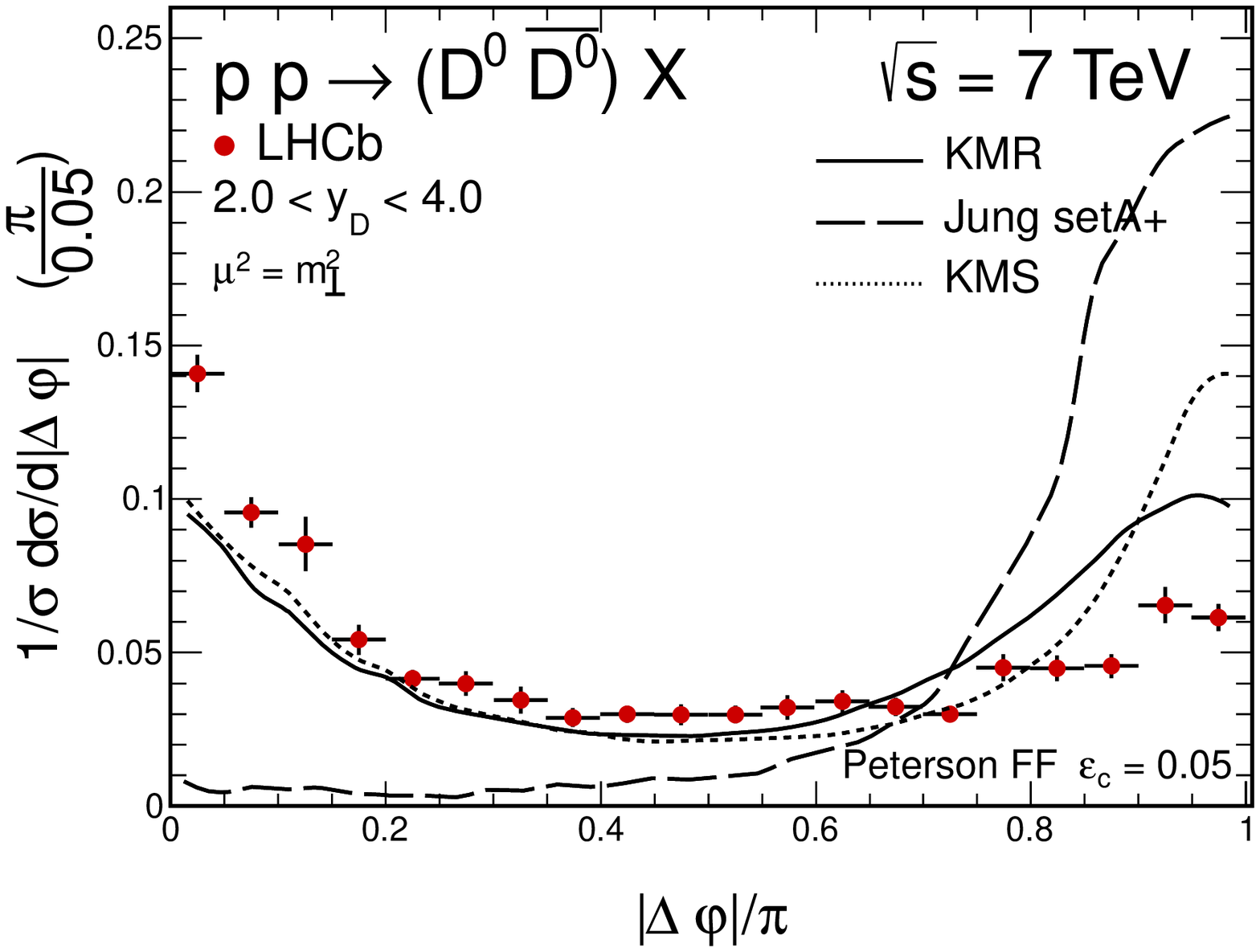}}
\end{minipage}
   \caption{
\small Invariant mass distribution of the $D^0 \bar D^0$ system (left) and distribution in relative azimuthal angle between $D^0$ and 
$\overline D^0$ (right) for different UGDFs.
}
 \label{fig:minv-lhcb-DDbar-2}
\end{figure}
The LHCb collaboration presented the distribution of the $D^0 \overline D^0$
invariant mass $M_{D^0 \overline D^0}$. In the left panel of Fig.~\ref{fig:minv-lhcb-DDbar-2} we show the
corresponding theoretical result for different UGDFs. Both, the KMR and KMS UGDFs provide the
right shape of the distribution. The dip at small invariant masses
is due to specific LHCb cuts on kinematical variables. 

The LHCb detector has almost full coverage in azimuthal angle.
In the right panel of Fig.~\ref{fig:minv-lhcb-DDbar-2} we show distribution in azimuthal
angle between the $D^0$ and $\overline D^0$ mesons $\varphi_{D^0 \overline D^0}$.
Both, the KMR and KMS UGDFs give the enhancement of the cross section
at $\phi_{D \bar D} \sim$ 0. This is due to the fact that these
approaches effectively include gluon splitting contribution, not included
in the case of the Jung UGDFs. However, still one can observe some small missing strength at small angles. It may suggest that within the KMR and KMS models the gluon splitting contribution is not fully reconstructed.

\section{Double charm production and meson-meson correlations}

Production of $c \bar c c \bar c$ four-parton final state is particularly
interesting especially in the context of experiments being carried out at the LHC
and has been recently carefully discussed \cite{Luszczak:2011zp,Maciula:2013kd}. 
The double-parton scattering formalism in the simplest form assumes two
independent standard single-parton scatterings. Then in a simple probabilistic picture, in the so-called factorized Ansatz, the differential cross section for DPS production of $c \bar c c \bar c$ system within the $k_{\perp}$-factorization approach can be written as:
\begin{eqnarray}
\frac{d \sigma^{DPS}(p p \to c \bar c c \bar c X)}{d y_1 d y_2 d^2 p_{1,t} d^2 p_{2,t} 
d y_3 d y_4 d^2 p_{3,t} d^2 p_{4,t}} = \nonumber \;\;\;\;\;\;\;\;\;\;\;\;\;\;\;\;\;\;\;\;\;\;\;\;\;\;\;\;\;\;\;\;\;\;\;\;\;\;\;\;\;\;\;\;\;\;\;\;\;\; \\ 
\frac{1}{2 \sigma_{eff}} \cdot
\frac{d \sigma^{SPS}(p p \to c \bar c X_1)}{d y_1 d y_2 d^2 p_{1,t} d^2 p_{2,t}}
\cdot
\frac{d \sigma^{SPS}(p p \to c \bar c X_2)}{d y_3 d y_4 d^2 p_{3,t} d^2 p_{4,t}}.
\end{eqnarray}

When integrating over kinematical variables one obtains
\begin{equation}
\sigma^{DPS}(p p \to c \bar c c \bar c X) = \frac{1}{2 \sigma_{eff}}
\sigma^{SPS}(p p \to c \bar c X_1) \cdot \sigma^{SPS}(p p \to c \bar c X_2).
\label{basic_formula}
\end{equation}
These formulae assume that the two partonic subprocesses are not correlated one with each
other and do not interfere. The parameter $\sigma_{eff}$ in the denominator of above formulae
from a phenomenological point of view is a non-perturbative quantity
related to the transverse size of the hadrons and has the dimension of a cross section.
The dependence of $\sigma_{eff}$ on the total energy at fixed scales is rather small and
it is believed, that the value should be equal to the total non-diffractive cross
section, if the hard-scatterings are really uncorrelated.
More details of the theoretical framework for DPS mechanism applied here can be found in Ref.~\cite{Maciula:2013kd}.

In turn, the elementary cross section for the SPS mechanism of double $c \bar c$
production has the following generic form:
\begin{equation}
d\hat{\sigma} = \frac{1}{2\hat{s}} \; \overline{|{\cal M}_{g g \rightarrow c \bar{c} c \bar{c}}|^2} \; d^{4} PS .
\end{equation}
where
\begin{equation}
d^{4} PS = \frac{d^3 p_1}{E_1 (2 \pi)^3} \frac{d^3 p_2}{E_2 (2 \pi)^3}
           \frac{d^3 p_3}{E_3 (2 \pi)^3} \frac{d^3 p_4}{E_4 (2 \pi)^3}
           \delta^4 \left( p_1 + p_2 + p_3 + p_4 - k_1 - k_2 \right) \; 
\end{equation}
is the 4-particle Lorentz invariant phase space, $k_1$ and $k_2$ are four-momenta of incoming gluons and $p_1, p_2, p_3, p_4$ are four-momenta of final charm quarks and antiquarks.

Neglecting small electroweak corrections and taking into account also $q \bar q$ annihilation terms, the hadronic cross section 
takes the following form:
\begin{eqnarray}
d \sigma &=& \int d x_1 d x_2 [
           g(x_1,\mu_F^2) g(x_2,\mu_F^2) 
           \; d \sigma_{gg \to c \bar c c \bar c} \nonumber \\
         &+& \Sigma_f \; q_f(x_1,\mu_F^2) \bar q_f(x_2,\mu_F^2) 
           \; d \sigma_{q \bar q \to c \bar c c \bar c}]
      \; .
\label{hadronic cross section}
\end{eqnarray} 

The matrix elements for single-parton scattering
were calculated using color-connected helicity amplitudes. They allow for an
explicit exact sum over colors, while the sum over helicities can be
done by using Monte Carlo methods. The color-connected amplitudes
were calculated following a recursive numerical Dyson-Schwinger approach.
More details about the SPS calculation and useful references can be found in Ref.~\cite{Hameren2014}.

\begin{figure}[!h]
\begin{minipage}{0.47\textwidth}
 \centerline{\includegraphics[width=1.0\textwidth]{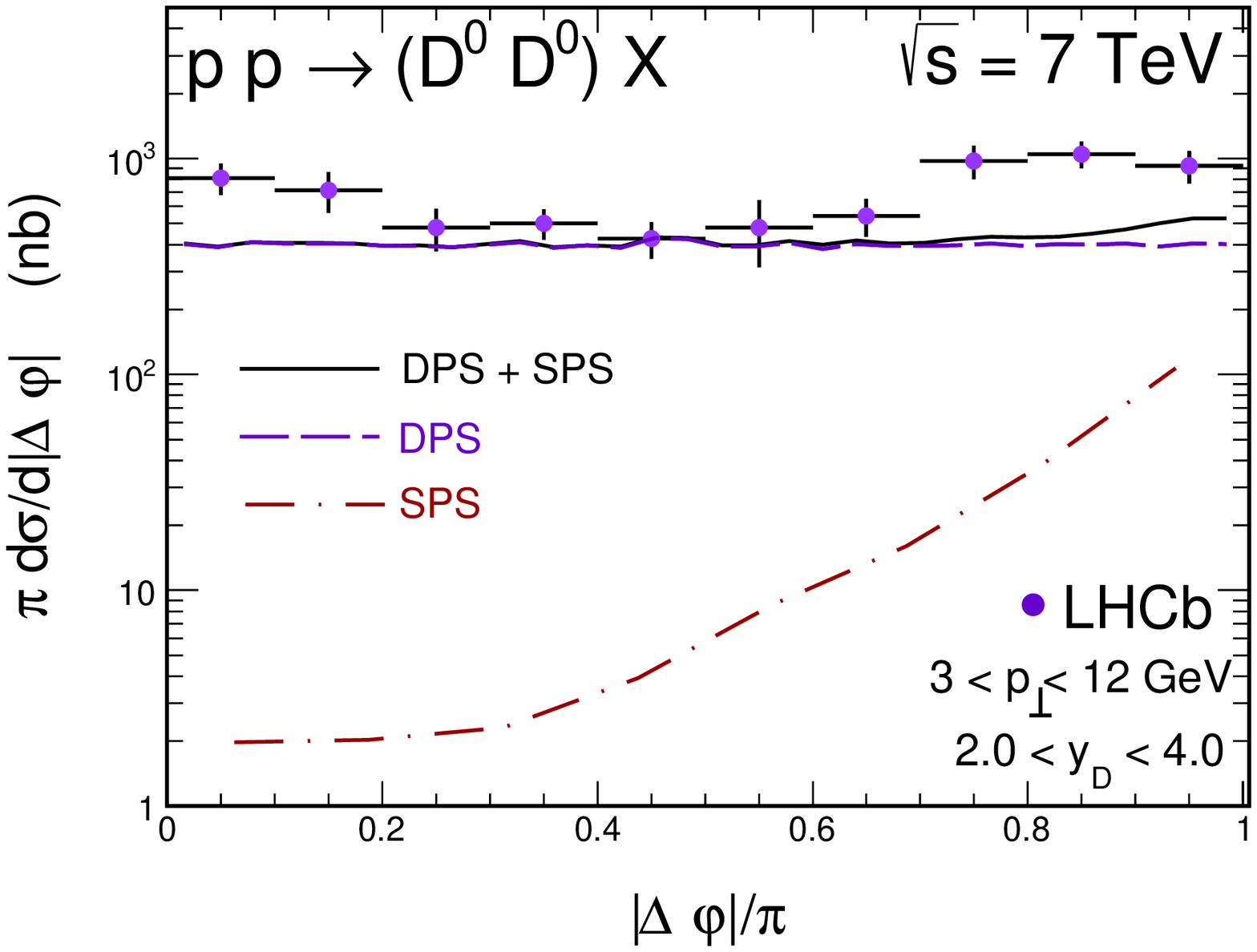}}
\end{minipage}
\hspace{0.5cm}
\begin{minipage}{0.47\textwidth}
 \centerline{\includegraphics[width=1.0\textwidth]{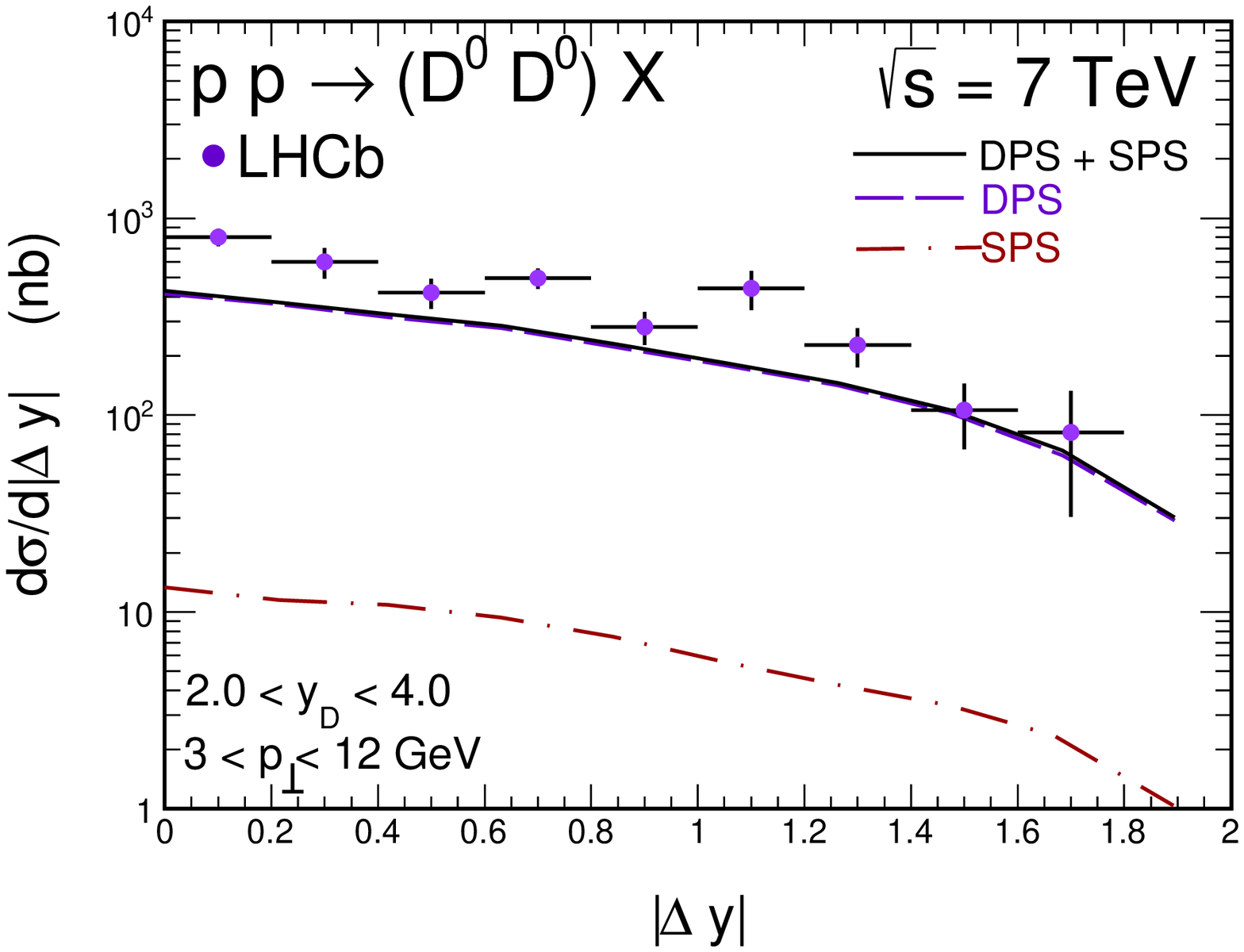}}
\end{minipage}

   \caption{
\small Azimuthal angle correlation between $D^0 D^0$ (left)
and distribution in rapidity difference between two $D^0$ mesons
(right) for DPS and SPS contributions.
}
 \label{fig:Phid-mesons}
\end{figure}

In Fig.~\ref{fig:Phid-mesons} we show azimuthal angle correlation (left panel) and distributions in relative
rapidity distance between two $D^0$ mesons (right panel) with kinematical cuts 
(rapidities and transverse momenta) corresponding to the LHCb
experiment. The shapes of the distributions are rather well reproduced.

\begin{figure}[!h]
\begin{minipage}{0.47\textwidth}
 \centerline{\includegraphics[width=1.0\textwidth]{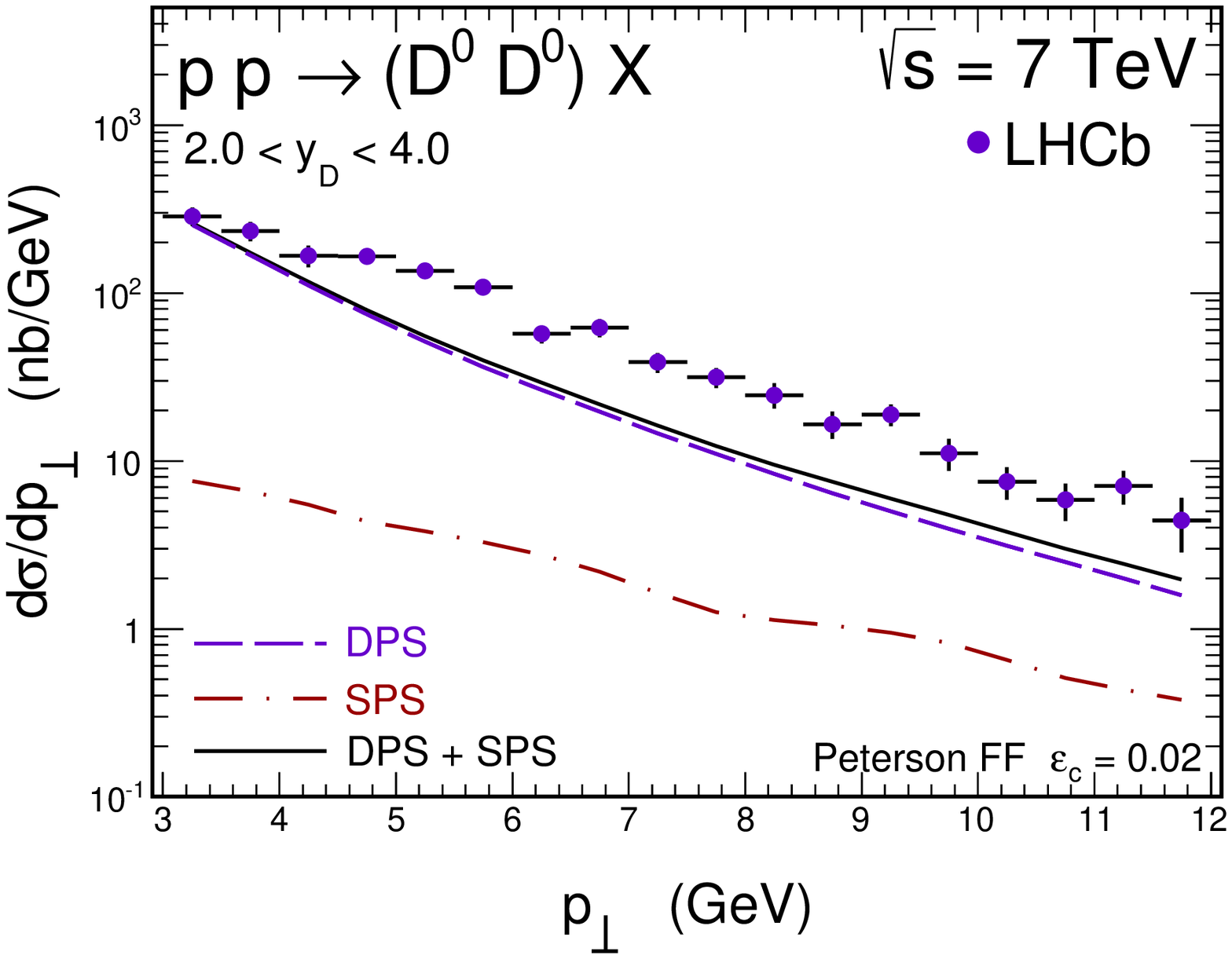}}
\end{minipage}
\hspace{0.5cm}
\begin{minipage}{0.47\textwidth}
 \centerline{\includegraphics[width=1.0\textwidth]{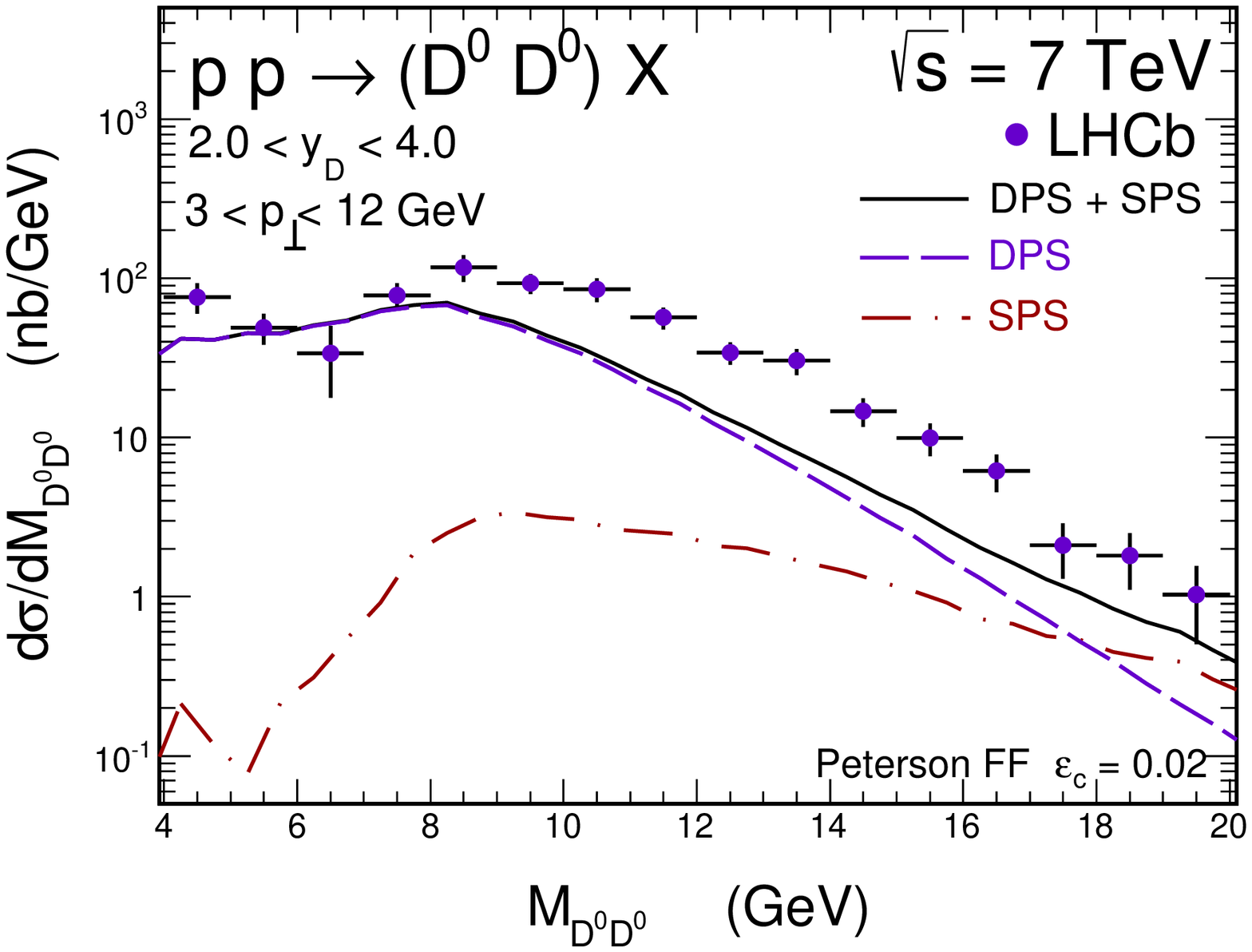}}
\end{minipage}

   \caption{
\small Distributions in meson transverse momentum when both mesons are
measured within the LHCb acceptance (left) and corresponding distribution
in meson invariant mass (right) for DPS and SPS contributions.
}
 \label{fig:Minv-and-pt-mesons}
\end{figure}

Other distributions in meson transverse momentum and two-meson invariant
mass are shown in Fig.~\ref{fig:Minv-and-pt-mesons}.
The shape in the transverse momentum is almost correct but some cross
section is lacking. Two-meson invariant mass distribution is shown in the right panel.
One can see some lacking strength at large invariant masses.

In the figures shown in this section the SPS contribution (dash-dotted line) is compared to the DPS
contribution (dashed line). The dominance of the DPS mechanism in description of the LHCb double charm data
is clearly confirmed. The DPS mechanism gives a sensible clarification of the measured
distribution, however some strength is still missing. This can be due to
3 $\to$ 4 processes discussed recently e.g. in Ref.~\cite{3to4-Gaunt}. This will be a subject 
of separate studies.
 
\end{document}